\documentclass[runningheads]{llncs}

\usepackage{times}
\usepackage{amsfonts} 
\usepackage{amsmath} 
\usepackage{amssymb} 
\usepackage[all]{xy} 

\begin{document}

\title{Efficient Certificateless Signcryption Tag-KEMs for Resource-constrained Devices}

\author{Wenhao Liu\inst{1}, Maurizio Adriano Strangio\inst{2} and Shengbao Wang\inst{1}}

\institute{Hangzhou Normal University, \\
School of Information Sciences and Engineering, Hangzhou, Zhejiang, China\\
\and
University of Rome ``Roma Tre", \\
Department of Mathematics and Physics, Rome, Italy\\}

\newcommand{\fracnoline}[2]{\begin{array}{c}{#1}\\{#2}\end{array}}
\newcommand{\assign}[2]{{#1}\stackrel{}{\leftarrow}{#2}}
\newcommand{\passign}[2]{{#1}\stackrel{R}{\leftarrow}{#2}}
\newcommand{\dpassign}[3]{{#1}\stackrel{#2}{\leftarrow}{#3}}
\newcommand{\tabbegin}[0]{\begin{tabular}{p{0.5cm}p{1.5cm}p{9.5cm}}}
\newcommand{\tabend}[0]{\end{tabular}}
\newcommand{\fracnl}[2]{\begin{center}\begin{tabular}{c}{#1}\\{#2}\end{tabular}\end{center}}

\newtheorem{Def}{Definition}
\newtheorem{Th}{Theorem} 
\newtheorem{Assump}{Assumption}

\maketitle

\begin{abstract}
Efficient certificateless one-pass session key establishment protocols can be constructed from key encapsulation mechanisms (KEMs) by making use of tags and signcryption schemes. The resulting primitives are referred to as Certificateless Signcryption Tag Key Encapsulation Mechanisms (CLSC-TKEMs). 

In this paper we propose two novel CLSC-TKEM protocols, the first, named LSW-CLSC-TKEM, makes use of the signature scheme of Liu et al., the second, named DKTUTS-CLSC-TKEM, is based on the  direct key transport using a timestamp (DKTUTS) protocol first described by Zheng. In order to achieve greater efficiency both schemes are instantiated on elliptic curves without making use of pairings and are therefore good candidates for deployment on resource constrained devices. 
\end{abstract}

\section{Introduction}

\label{sec:INTRO}

Certificateless cryptography (CLC), introduced by Al-Riyami and Paterson \cite{RIPA03}, does not require a public-key infrastructure (PKI) for digital certificate management and does not suffer from the inherent key escrow feature of identity-based cyptography (IBC). 
A certificateless scheme continues to make use of a trusted third party known as the key generating center (KGC) which, as opposed to IBC, does not have access to the user's private key. In the CLC setting user private keys are constructed from two partial secrets: one generated by the KGC computed from the user's identity and a secret master key and a secret value chosen by the user itself.  The scheme is not identity-based, because the public key is no longer exclusively computable from a user's identity. When Alice wants to send a message to Bob using a certificateless scheme, she must obtain Bob's public key. However, no authentication of Bob's public key is necessary and no certificate is required (as normally would be the case with a PKI). 

The generation of a cryptographic secret key and its encryption with a public key encryption scheme is generally known as key encapsulation mechanism (KEM). Further encrypting a message with the secret key and a symmetric key encryption scheme is known as a data encryption mechanism (DEM). In general, the resulting KEM-DEM schemes combine advantages of both symmetric and asymmetric cryptographic techniques thus giving rise to secure and efficient hybrid public key encryption schemes \cite{HHK06,DENT02}. Efficient one-pass session key establishment protocols \cite{GBG07} can be constructed based on KEMs by making use of tags \cite{AGKS05} and signcryption schemes \cite{ZHENG97} appropriately instantiated in the CLC setting. The resulting primitives are referred to as Certificateless Signcryption Tag Key Encapsulation Mechanisms (CLSC-TKEMs). The use of a signcryption scheme implies additional important security properties such as user (sender) non-repudiation which are derived from the use of a digital signature scheme. 

In this paper we propose two CLSC-TKEM protocols, the first, named LSW-CLSC-TKEM, makes use of the signature scheme of Liu et al. \cite{LXWHH14}, the second, named DKTUTS-CLSC-TKEM, is based on the  direct key transport using a timestamp (DKTUTS) protocol described in \cite{ZHENG98}. In order to achieve greater efficiency both schemes can be instantiated on elliptic curves without making use of pairings and are therefore ideal candidates for deployment on resource constrained devices. 

\section{Related work}

In recent work, Jongho Won \emph{et al.} \cite{JSB15} proposed an efficient CLSC-TKEM protocol (eCLSC-TKEM) for securing communications between drones and smart objects. According to the authors, the protocol supports authenticated key agreement, non-repudiation, and user revocation and significantly reduces the time required to establish a shared key between a drone and a smart object by minimizing the computational overhead on the smart object (since the protocol does not make use of pairings). 
A problem with this protocol is possibly due to the user revocation technique introduced by the authors which allows expiration of user partial private keys and therefore requires subsequent reissue of a key and distribution to the user by the KGC.   
Much like the aforementioned protocol, the CLSC-TKEM scheme of Seo \emph{et al.}\cite{SB13} is also inefficient from the computational perspective if the target recipient is a low-power resource-limited device. 

In \cite{TS15} the authors propose a generic architecture for a CLSC-TKEM that is based on a true random number generator (TRNG) to produce secure cryptographic secret keys for a KEM/DEM scheme.

\section{Theoretical Framework for Certificateless Signcryption Tag-KEMs}
\label{sec:SCHEME}

We refer to the framework of Signcryption Tag-KEMs (SC-TKEM) introduced by Bjorstad and Dent \cite{BJDE05} and extend it to the CLC setting.
A CLSC-TKEM is defined as the tuple of six algorithms described below:
\begin{enumerate}

\item \textbf{Setup}: A probabilistic common parameter generation algorithm that takes as input a security parameter $1^{k}$ and returns all the global system parameters $\Omega$ needed by users of the scheme, such as choice of groups or hash functions. The algorithm also outputs the private/public key pair ($sk_{KGC}, pk_{KGC}$) of the KGC.

\item \textbf{PartialPrivateKeyExtract}: A probabilistic key generation algorithm that takes as input the identity $ID_E$ of a generic entity $E$, $\Omega$ and outputs the  partial private key $d_E$ of $E$. This algorithm is generally run by the KGC which must thereafter deliver the partial private key $d_E$ to $E$ through a secure channel. 

\item \textbf{GenUserKeys}: A probabilistic key generation algorithm that takes as input $\Omega$ and generates the private/public key pair ($x_E, P_E$) of entity $E$. Entity $E$ sets $sk_E= (x_E,d_E)$ as its full private key. 

\item \textbf{SymmetricKeyGen}: A probabilistic symmetric key generation algorithm that takes as input the public key $pk_B$ of the recipient  entity $B$ and outputs the symmetric key $K$ and internal state information $\omega$. 

\item \textbf{Encapsulation}: A probabilistic key encapsulation algorithm that receives as input the state information $\omega$, an arbitrary tag $\tau$, the full private key $sk_A$ of the sender $A$ and returns an encapsulation  $\phi$.

\item \textbf{Decapsulation}:  A deterministic decapsulation/verification algorithm that takes as input the public key $pk_A$ of the sender $A$, the full private key $sk_B$ of the recipient, an encapsulation $\phi$  and a tag $\tau$ and returns either the symmetric key $K$ or the  unique error symbol $\perp$.

For the CLSC-TKEM to be sound, the decapsulation/verification algorithm must return the correct key $K$ whenever the encapsulation $\phi$ is correctly formed and the corresponding keys and tag are supplied.

\end{enumerate}

\section{The LSW-CLSC-TKEM protocol specification}
\label{sec:CLSCTKEM} 

The LSW-CLSC-TKEM protocol,  based on the signature scheme of Liu et al. \cite{LXWHH14}, 
is completely specified by the six polynomial time algorithms specified below:
\begin{enumerate}
\item \textbf{Setup}: On input the security parameter $k\in \mathbb{Z}+$, the KGC returns the system parameters $\Omega$ (see below) and the KGC's master private key $x_{msk}$. The KGC also performs the following steps:
\begin{itemize}
\item Chooses a $k$-bit prime $q$, generates a cyclic additive group $G$, a cyclic multiplicative group $G_2$ both of order $q$ and defines the tuple $\langle F_q,E/F_q,G,G_2,P\rangle$, with $P$ generator of $G$.
\item Chooses the master key $x_{msk} \in_R \mathbb{Z}^*_q$ uniformly at random and computes the system public key $P_{pub} = x_{msk}P$.
\item Chooses the cryptographic hash functions $H_{1}: \{0, 1\}^* \times G  \rightarrow \mathbb{Z}^*_q$,  $H_{2}: \{0, 1\}^* \rightarrow \mathbb{Z}^*_q$ and $H_{3}: \{0, 1\}^*  \rightarrow \{0, 1\}^*$;
\item Publishes the global system parameters $\Omega = \langle F_q,E/F_q,G,G_2,P,P_{pub},H_1,H_2\rangle$.
\end{itemize} 

\item \textbf{PartialPrivateKeyExtract}: For entity $A$, with identity $ID_A$, the KGC chooses $r_A \in_R \mathbb{Z}^*_q$  computes $R_A=r_A P$, $h_A=H_1(ID_A,R_A), d_A=r_A + x_{msk} h_A \text{ mod } q$ and delivers the  partial private key  $d_A$ to user $ID_A$ through a secret channel. Entity A can validate her key by verifying that $d_A P=R_A + h_A P_{pub}$.

\item \textbf{GenUserKeys}: Entity A with an identity $ID_A$ chooses $x_A \in_R \mathbb{Z}^*_q$ as its secret value and generates the corresponding public key $P_A = x_AP$. Furthermore, entity A sets $sk_A = (x_A,d_A)$ as its full private key and $pk_A = (P_A,R_A)$ as its full public key. 

\item \textbf{SymmetricKeyGen}: Given the sender identity $ID_A$, the receiver identity $ID_B$ and the full public key $pk_B$ as inputs, entity A (the sender) proceeds as follows: 
\begin{itemize}
\item Chooses $u_A \in_R \mathbb{Z}_q^*$ and computes $U = u_A(R_B+H_1(ID_B,R_B)P_{pub}+P_B)$;
\item Computes $X = u_AP$ and $K = H_1(X,U,ID_A,ID_B)$;
\item Outputs $K$ and $\omega = (u_A,ID_A,sk_A,ID_B,pk_B,X,U)$.
\end{itemize}

\item \textbf{Encapsulation}: On input $\omega$, an arbitrary tag $\tau$, the full private key $sk_A$, entity $A$ obtains the encapsulation $\phi$ by performing the following operations:
\begin{itemize}
\item Selects $a \in_R \mathbb{Z}_q^*$ and computes $Q=aP$;
\item Computes $h = H_2(\tau,ID_A,ID_B,R_A,R_B,P_A,P_B,Q,X,U)$;
\item Computes $s=a/(hx_A+d_A)$;
\item Sets $\sigma=(s,h)$  and outputs $\phi = \langle Q,U,\sigma\rangle$.
\end{itemize}

\item \textbf{Decapsulation}: On input the encapsulation $\phi$, tag $\tau$, the sender's identity $ID_A$, full public key $pk_A$, the receiver's identity $ID_B$ and the full private key $sk_B$, the recipient entity $B$ performs the following operations:
\begin{itemize}
\item Computes $(d_B+x_B)^{-1}\cdot U = u_AP = X$;
\item Computes $h=H_2(\tau,ID_A,ID_B,R_A,R_B,P_A,P_B,Q,X,U)$;
\item If $s(hP_A+R_A+H_1(ID_A,R_A)P_{pub}) \ne Q$, returns with an invalid encapsulation error $\perp$;
\item Otherwise, accepts the key $K = H_1(X,U,ID_A,ID_B)$.
\end{itemize}
The correctness of the protocol is determined as follows:
\begin{eqnarray*} 
s(hP_A+R_A+H_1(ID_A,R_A)P_{pub}) & = & a(hx_A+d_A)^{-1}(hP_A+R_A+h_AP_{pub}) \\ 
& = & a(hx_A+d_A)^{-1}(hx_AP+r_AP+h_AxP) \\
& = & a(hx_A+d_A)^{-1}(hx_A+r_A+h_Ax)P \\
& = & a(hx_A+d_A)^{-1}(hx_A+d_A)P \\
& = & aP = Q\\
\end{eqnarray*}  
\end{enumerate}

\section{The DKTUTS-CLSC-TKEM protocol specification}
\label{sec:DKTUTSCLSCTKEM} 

The DKTUTS-CLSC-TKEM protocol, based on the direct key transport using a timestamp (DKTUTS) protocol described in \cite{ZHENG98}, is completely specified by the six polynomial time algorithms specified below:

\begin{enumerate}
\item \textbf{Setup}: On input the security parameter $k\in \mathbb{Z}+$, the KGC returns two system parameters: $\Omega$ and the KGC's master private key $x_{msk}$. The KGC also performs the following steps:
\begin{itemize}
\item Chooses a $k$-bit prime $q$, generates a cyclic additive group $G$, a cyclic multiplicative group $G_2$ both of order $q$ and determines the tuple $\langle F_q,E/F_q,G,G_2,P\rangle$, with $P$ generator of $G$.
\item Chooses the master key $x_{msk} \in_R \mathbb{Z}_q*$ uniformly at random and computes the system public key $P_{pub} = x_{msk}P$.
\item Chooses the cryptographic hash functions $H_{1}: \{0, 1\}^* \times G  \rightarrow \mathbb{Z}^*_q$,  $H_{2}: \mathbb{Z}_q^* \rightarrow \{0, 1\}^*$ and a keyed hash function $F_K: \{0, 1\}^*  \rightarrow \{0, 1\}^*$;
\item Chooses the symmetric encryption scheme ($E_K(\cdot),D_K(\cdot)$);
\item Publishes the global system parameters $\Omega = \langle F_q,E/F_q,G,G_2,P,P_{pub},H_1,H_2,KH,E,D\rangle$.
\end{itemize} 

\item \textbf{PartialPrivateKeyExtract}: For entity $A$, with identity $ID_A$, the KGC chooses $r_A \in_R \mathbb{Z}^*_q$  computes $R_A=r_A P$, $h_A=H_1(ID_A,R_A), d_A=r_A + x_{msk} h_A \text{ mod } q$ and delivers the  partial private key  $d_A$ to user $ID_A$ through a secret channel. Entity A can validate her key by verifying that $d_A P=R_A + h_A P_{pub}$.

\item \textbf{GenUserKeys}: Entity A with an identity $ID_A$ chooses $x_A \in_R \mathbb{Z}^*_q$ as its secret value and generates the corresponding public key as $P_A = x_AP$. Furthermore, entity A sets $sk_A = (x_A,d_A)$ as its full private key and $pk_A = (P_A,R_A)$ as its full public key. 

\item \textbf{SymmetricKeyGen}: Given the sender identity $ID_A$, the receiver identity $ID_B$ and the full public key $pk_B$ as input, the sender proceeds as follows: 
\begin{itemize}
\item Chooses $K \in_R \{0,1\}^{l_k}$ and $x,a \in_R \mathbb{Z}^*_q$;
\item Computes $U=aP$ and $X=x(R_B+H_1(ID_B,R_B)P_{pub}+P_B)$;
\item Computes $(k_1,k_2)= H_2(X+U)$;
\item Outputs $K$ and $\omega = (x,k_1,k_2,TS,ID_A,ID_B,pk_B,X,U)$ where $TS$ is a suitably defined timestamp.
\end{itemize}

\item \textbf{Encapsulation}: On input $\omega$, an arbitrary tag $\tau$, the full private key $sk_A$, entity $A$ obtains the encapsulation $\phi$ by performing the following computations:
\begin{itemize}
\item Computes $c=E_{k_1}(K,TS,\tau,ID_A,ID_B,R_A,R_B,P_A,P_B,X,U)$, \\ $r = F_{k_2}(K,TS,\tau,ID_A,ID_B,R_A,R_B,P_A,P_B,X,U)$ \\ and $s=x/(r+x_A) \text{ mod q}$; 
\item Outputs $\phi = \langle U,c,r,s \rangle$.
\end{itemize}

\item \textbf{Decapsulation}: On input the encapsulation $\phi$, tag $\tau$, the sender's identity $ID_A$, full public key $pk_A$, the receiver's identity $ID_B$ and the full private key $sk_B$, the recipient entity $B$ performs the following computations:
\begin{itemize}
\item Computes $X'=s(d_B + x_B)(P_A + rP)$ and $H_2(X'+U)=(k_1,k_2)$;
\item Computes $K,TS,\tau,ID_A,ID_B,R_A,R_B,P_A,P_B,X,U' = D_{k_1}(c)$ and \\ $r' = F_{k_2}(K,TS,\tau,ID_A,ID_B,R_A,R_B,P_A,P_B,X,U')$;
\item If $TS$ is not fresh or $U \ne U'$ or $X' \ne X$ or $r' \ne r$, returns with an invalid encapsulation error $\perp$;
\item Otherwise, accepts the key $K$.
\end{itemize}
The correctness of the protocol is determined as follows:
\begin{eqnarray*} 
s(d_B+x_B)(P_A+rP) & = & sd_B(P_A+rP) + sx_B(P_A+rP) \\ 
& = & s(x_Ad_BP+rd_BP) + s(x_Ax_BP + rx_BP) \\
& = & s(x_A+r)d_BP + s(x_A+r)P_B \\
& = & xd_BP+xP_B\\
& = & x(d_BP + P_B) = X\\
\end{eqnarray*}  
\end{enumerate}

\section{On the efficiency and security of CLSC-TKEM protocols}

In this section we compare four CLSC-TKEM protocols from two perspectives: computational load and security properties.
Tables \ref{TAB:COMPS} and \ref{TAB:COMPR} summarize the computational cost of the sender and recipient principals respectively. The features that are taken into account are: a) online and offline exponentiations, the former refer to the operations that are performed during running instances of the protocols while the later consider the pre-computation of values that can be performed before protocol execution (this data must be safely stored in the sender device); b) field inversions (fld inv.); c) field multiplications (fld mult.) and d) decryption operations with a symmetric cipher. 

\begin{table*}
  \centering
\caption{Computational efficiency of CLSC-TKEM protocols - sender}
\begin{tabular}{|c|c|c|c|c|c|}
  \hline
  \emph{Protocol} & \emph{online exp.} & \emph{offline exp.} & \emph{fld inv.} &  \emph{fld mult.} &  \emph{encryption}\\
  \hline
  \hline
  CLSC-TKEM\cite{SB13} & 2EM & 0EM & 0 & 2 & 0 \\ 
  eCLSC-TKEM\cite{JSB15} & 4EM & 2EM & 0 & 0 & 0 \\
  LSW-CLSC-TKEM  & 3EM & 0EM & 1 & 2 & 0  \\
  DKTUTS-CLSC-TKEM & 2EM & 0EM & 1 & 1 & 1  \\
  \hline
\end{tabular}
\label{TAB:COMPS}
\end{table*}

\begin{table*}
  \centering
\caption{Computational efficiency of CLSC-TKEM protocols - recipient}
\begin{tabular}{|c|c|c|c|c|c|}
  \hline
  \emph{Protocol} & \emph{online exp.} & \emph{offline exp.} & \emph{fld inv.} &  \emph{fld mult.} &  \emph{decryption}\\
  \hline
  \hline
  CLSC-TKEM\cite{SB13} & 5EM & 3EM & 0 & 0 & 0 \\ 
  eCLSC-TKEM\cite{JSB15}  & 4EM & 2EM & 0 & 0 & 0 \\
  LSW-CLSC-TKEM  & 3EM & 0EM & 1 & 0 & 0  \\
  DKTUTS-CLSC-TKEM & 2EM & 0EM & 0 & 1 & 1  \\
  \hline
\end{tabular}
\label{TAB:COMPR}
\end{table*}

Table \ref{TAB:SEC} summarizes the security properties of the same CLSC-TKEM protocols considered above. The security properties that are taken into account are: a) sender partial forward secrecy (sPFS); b) user authentication (sender); c) non repudiation (sender); d) user revocation; e) security proof, indicates whether a formal security proof exists for the protocol.  

All protocols considered in table \ref{TAB:SEC} do not guarantee \emph{forward secrecy} (FS). For these typical one-pass key transport schemes where the recipient does not contribute to the computation of the session key the appropriate notion is that of \emph{partial forward secrecy} (PFS) i.e. if compromise of the long-term keys of one or more specific principals does not compromise the session keys established in previous protocol runs involving those principals \cite{BM03}. In particular, for the protocols we are discussing it makes sense to consider \emph{sender partial forward secrecy} \cite{GBG07} (respect to a passive adversary that can corrupt peers to obtain long-term keying material such as the private key and that does not modify protocol messages in transit through the network). 

\begin{table*}
  \centering
\caption{Security properties of CLSC-TKEM protocols}
\begin{tabular}{|c|c|c|c|c|c|c|}
  \hline
  \emph{Protocol} & \emph{sPFS} & \emph{user auth.} &  \emph{non-repud.} &  \emph{user revoc.} & \emph{sec. proof}\\
  \hline
  \hline
  CLSC-TKEM\cite{SB13}   	& yes & yes & yes & no  & yes\\  
  eCLSC-TKEM\cite{JSB15} 	& yes & yes & yes & yes & yes\\
  LSW-CLSC-TKEM  					& yes & yes & yes & no  & no \\
  DKTUTS-CLSC-TKEM 				& yes & yes & yes & no  & no \\
  \hline
\end{tabular}
\label{TAB:SEC}
\end{table*}

\section{Final remarks}

In this paper we have addressed the problem, introduced by \cite{JSB15}, of ensuring that drones can perform secure communications with many different smart objects, such as sensors and embedded devices. The authors propose a Certificateless Signcryption Tag Key Encapsulation Mechanism (eCLSC-TKEM) that minimizes the computational load on the receiving resource-constrained mobile device. We have proposed two constructions that achieve better performance in terms of the computational overhead required by the recipient.  

However, resource-constrained devices are often more susceptible to private key exposure therefore forward secrecy (of the recipient) may be indeed a desirable security property for the above protocols. Depending on the target application, when forward secrecy is necessary a possible option is to employ two-pass key agreement protocols at the expense of a greater computational cost for the recipient (the protocols described in this paper can be modified into equivalent key agreement versions). Another possibility of mitigating the consequences of user corruption by an adversary is to use a key evolving mechanism so that keys are updated periodically an thus damage is limited to the period of validity of the exposed key \cite{FRK06}. 

\bibliographystyle{abbrv}

\end{document}